# Spreadsheet Validation and Analysis through Content Visualization

Richard Brath, Michael Peters
Oculus Info, Inc.
2 Berkeley Street, Suite 600, Toronto, Canada, M5A 4J5
{richard.brath@oculusinfo.com, mike.peters@oculusinfo.com }

**ABSTRACT**

*Visualizing spreadsheet content provides analytic insight and visual validation of large amounts of spreadsheet data. Oculus Excel Visualizer is a point and click data visualization experiment which directly visualizes Excel data and re-users the layout and formatting already present in the spreadsheet.*

## 1. INTRODUCTION

Spreadsheets have been widely adopted as simple and powerful tools for free-form data manipulation. The free-form nature of a spreadsheet, however, can make management and analysis of a spreadsheet difficult.

Visualization is a powerful technique for identifying trends, patterns and anomalies in data sets and has been applied to spreadsheets in various ways. Oculus Excel Visualizer is a point and click visualization experiment which directly visualizes Excel data and re-uses the layouts and formats already present in the visualization. Excel Visualizer has been used for data validation and analysis on very large spreadsheets, including validation of data quality for financial time series data; validation of modeled results for financial forecasting data; and analysis of large pivot tables.

## 2. BACKGROUND

Visualization is a potentially powerful means for reviewing spreadsheets; and has been used with spreadsheets in many ways, including:

- **Conditional Formatting and Data Bars** are functionality within spreadsheets for highlight cells based on the cell values, for example, Excel Highlighter from Spheresoft and Microsoft’s new data bar feature [Gai05]. These simple techniques provide for some easy visual analysis.

- **Calculation dependencies** can be represented as graphs (nodes and links) to illustrate the underlying calculation network. In addition to built-in auditing functions within spreadsheets such as Excel, researchers such as [Iga99], [Cha01], [Shi99] and [Cle03] have extended these concepts.

- **Screening and Grouping** can be used to highlight groups of cells sharing common attributes, such as similar formulas, formats and values. Examples include Spreadsheet Detective (Southern Cross Software), OAK (Operis Group). Researchers such as [Saj00] and [Cle03] have extened these into the concept of semantic content analysis.

- **Charting and Data Visualization Functionality** associated with spreadsheets provide for some visual data analysis, but require the spreadsheet data to be re-organized into layout that faciliates the tool. These tools cannot utilize data outside the immediate cell range required by the tool and typically ignore any spreadsheet formatting. There are many commercially available tools in this category, such as DataDesk (DataDesk), Xcelsius (Business Objects), SeeIT and ScatterPlot (Visible Decisions), Advizor and Advizor 2000 (Advizor Solutions), Tableau (Tableau Software), and Spotfire (Spotfire U.S.).

- **Viewing Techniques** such as Magic Lenses [Bie93] fish eye views, hyperbolic distortions [Fur86] and zooming interfaces, such as Table Lens [Rao94] can be applied to directly visualizing spreadsheets. However, distortion techniques, such as a fish-eye view, can preserve both the spreadsheet data and formatting content but does not facilitate visual comparison of numeric values by graphically encoding those values with perceptual cues of height or color. The Table Lens technique does convert numeric values to graphical representations but implementations of this technique require consistency of data types across columns.

We have created a visualization called Oculus Excel Visualizer that enables direct visualization of a spreadsheet without the need for re-organzing the data nor requires the configuration of features such as screening or conditional formatting. The approach of Excel Visualizer is the WYSIWYG (what-you-see-is-what-you-get) paradigm originally coined during the evolution of early graphic user interfaces. This approach seeks to characterize representations of text documents and spreadsheets as closely as possible to their physical output, thereby enabling the user to reduce the cognitive effort between the two representations. In applying the WYSIWYG approach to spreadsheets we attempt to reduce the cognitive gap between the largely numerical and text based spreadsheet and its semantic, graphical representation. We have applied WYSIWYG to spreadsheets as follows:

- The visualization uses the same row and column cell structure for organizing the scene.
- Text data in the spreadsheet continues to be represented as text within cells in the visualization.
- Numeric data is represented as graphical objects (bars by default) in the visualization.
- Wherever possible all other spreadsheet formatting (cell background color, cell borders) are re-used within the visualization.

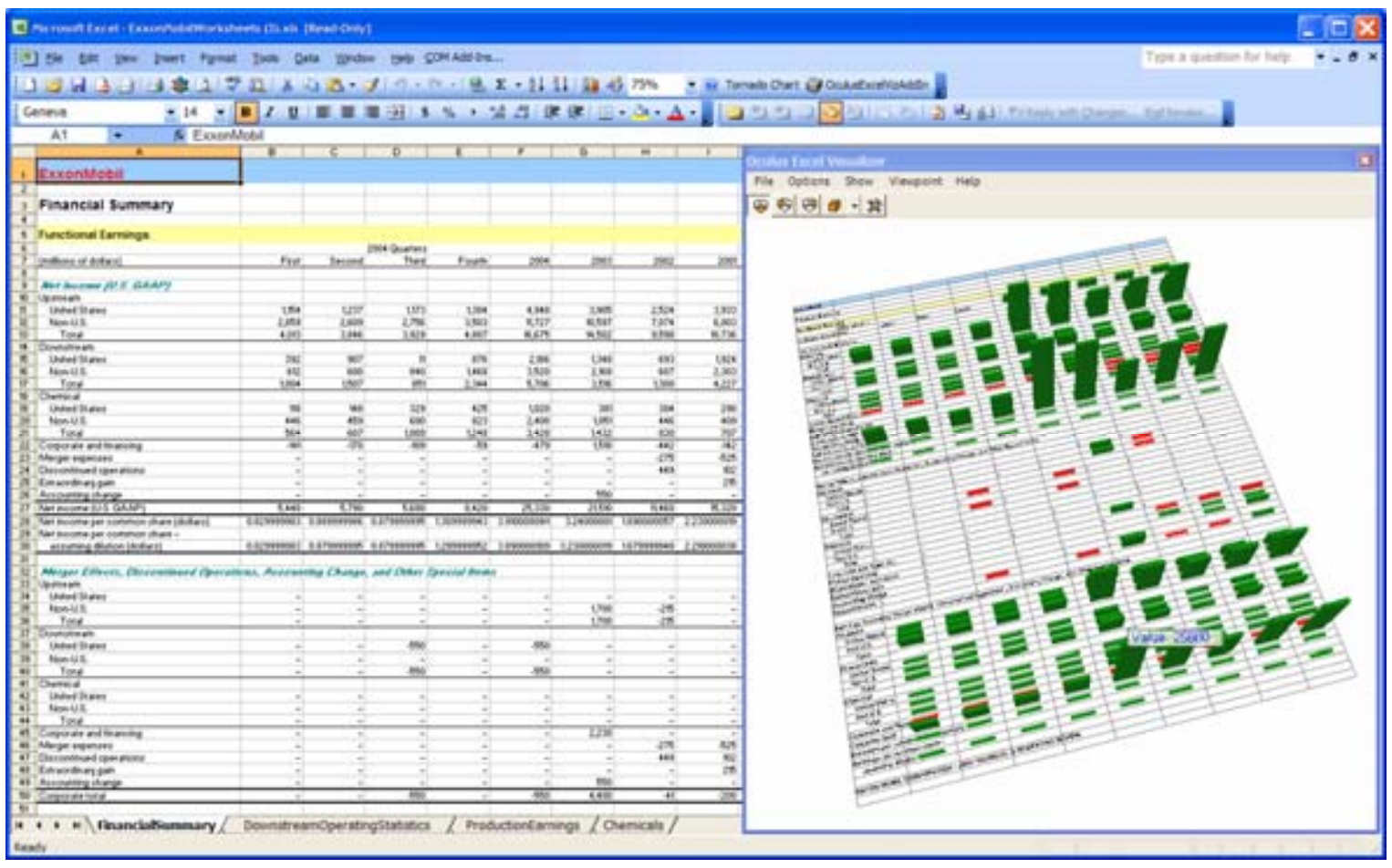

Figure 1. A simple spreadsheet from an annual report. On the left is an income statement with a variety of formatting and on the right is Excel Visualizer viewing the same data.

# 3. EXCEL VISUALIZER OVERVIEW

The WYSIWYG paradigm can be a very powerful tool for user cognition. In general, literal metaphors can be powerful user-interface techniques because of the ease of understanding. Both WYSIWYG and literal metaphors suffer from limitations as well: they can be difficult to extend beyond the core metaphor.

## 3.1 Key Features

Following the basic WYSIWYG paradigm provides several benefits for the user of the visualization, such as preservation of formatting and ease of data manipulation.

**Bi-Directional Synchronization.** Excel Visualizer provides synchronization between the spreadsheet and the visualization: if any cell in the spreadsheet is changed, then the visualization immediately updates. Clicking on a bar or cell in the visualization immediately selects and highlights the corresponding cell in the spreadsheet. The end-user can easily move back and forth between the two representations.

**3D Navigation.** The user can dynamically pan, rotate and zoom in and out of the scene. This provides a means to start with a high-level overview ("does this data seem right at a high level?") and then zoom in to inspect finer details as required. Combined with the bi-directional synchronization with the source spreadsheet, the user can investigate individual bars and see them within the context of the overall data in the spreadsheet..

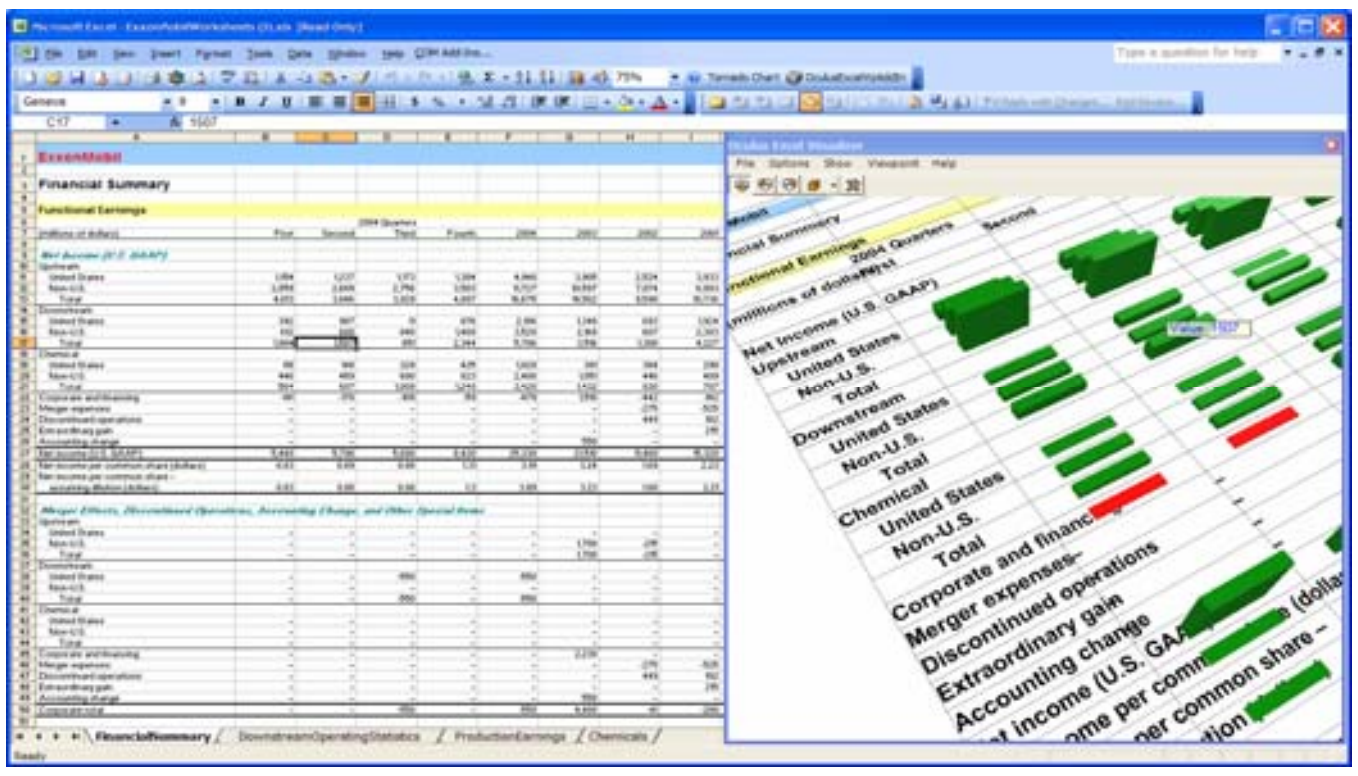

Figure 2. Excel Visualizer tightly zoomed in on a region of data. The user clicks on a bar which provides a tooltip of the numeric value and highlights the corresponding cell in the spreadsheet.

**Visual Scalability.** Excel Visualizer utilizes a powerful 3D rendering engine and can graphically display large amounts of data. Excel Visualizer can display in excess of 75,000 cells simultaneously in one screen. This would represent more than 30 screens worth of numeric data in a typical spreadsheet view. Viewing large amounts of data at once can help the user spot trends, patterns and outliers.

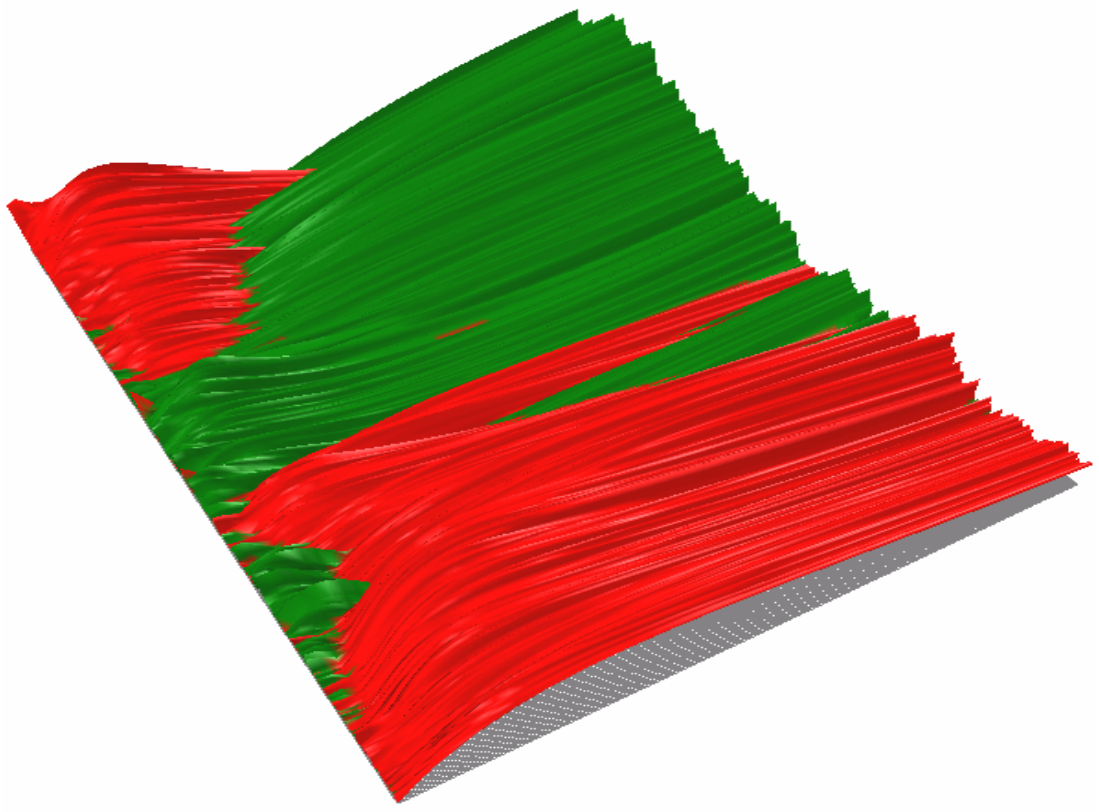

Figure 3. A view on 60 x 610 array of cells showing the difference between two financial instruments over a three year time period. Twists and dips are immediately visible.

**Format Preservation.** Spreadsheet users invest significant time and effort adjusting the fonts, colors, borders, and white space to enhance the readability and usability of spreadsheets. Through these formatting techniques, end-users leverage Gestalt principles such as proximity, similarity, figure and ground separation to organize the information. By maintaining the formatting, these extra layers of semantic value are available to the visualization user.

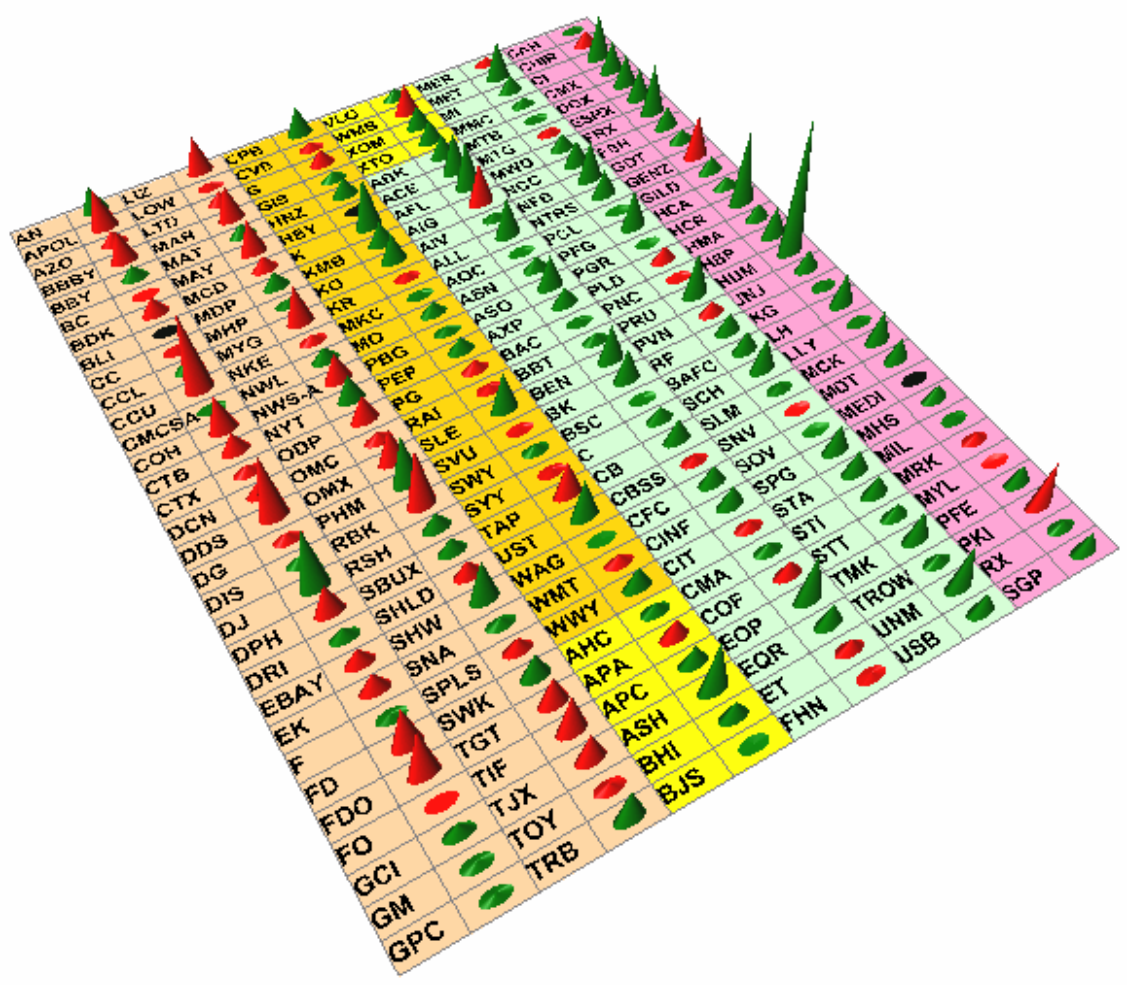

Figure 4. Cell shading, used to group objects together, appears in the Visualizer as well.

**Data Type Flexibility**. Generally, spreadsheet users have already formatted their data into specific data types – for example, date formats, currency formats are already set. With Excel Visualizer, there is no need for the user to adjust data types or go through a wizard. Instead, the application intelligently handles numeric, text and missing data, allowing the user much more freedom for ad hoc visual exploration.

**Data manipulation.** Users of visualizations for exploratory data analysis may find the need to edit data during the exploration. In many visualization software packages, this may require editing the data in another application (such as a spreadsheet) to perform operations such as correcting the value of an erroneous data point, creating a new derived data attribute (e.g. Profit equals Sales minus Costs), or computing sums or other statistics. Once the data has been modified, the steps to re-import the data may need to be repeated. The process can be tedious and error prone.

Spreadsheets can be quite powerful for free-form and ad hoc data manipulation and modeling, such as pivot tables, sorting, filtering, programming, etc. A WYSIWYG-based visualization tightly coupled to the spreadsheet provides the user with full access to their extremely powerful and familiar spreadsheet tools.

**Annotation.** The annotation of interesting results can be challenging in many visualization systems. Using the spreadsheet, the user can add annotations in adjacent cells, which then appear in the visualization.

**Visual flexibility.** Different datasets lend themselves to different visual representations. For example, continuous time series data may be better represented as a surface rather than a matrix of disparate bars.

### 3.2 Limitations

Following a strong literal metaphor does present some limitations. The most challenging limitations of the WYSIWYG metaphor for spreadsheet visualization include:

**Occluding Bars.** In a typical spreadsheet, labels are positioned at the top and to left of a range of cells. Bars representing data may easily occlude labels along the top.

**Labels.** When using Excel Visualizer to view a large spreadsheet area, text is essentially too small to be readable. However, Excel Visualizer can be used effectively to provide a "big picture" with the spreadsheet providing the "focus" for the region of interest that the user clicks on.

**Data Normalization.** All numeric values in the visualization are normalized across the entire range of selected data. This can make it difficult to compare different types of values, for examples, one range of cells in units of millions of dollars and another range of cells in units of cents. By default, Excel Visualizer normalizes all numeric data uniformly. This could be overcome by some degree by normalizing numeric data based on common data formats – e.g. all cells formatted as percentages are normalized separately from all cells formatted as currency. This would also require a visual cue to the user to indicate that there are two different numeric types in the visual display, perhaps using categorical signifier such as shape or color to indicate different numeric types.

**Fixed Cell Sizes**. Excel Visualizer represents the selected range of data in fixed column and row widths and heights. This presents a problem when columns are used for text strings that don't fit it the standard cell width.

**Multifunctioning Visual Elements**. Information visualization can be very valuable for analyzing multiple variables at once through the use of multifunctioning data elements – for example the location, size, orientation, shape, etc of various markers can convey different data attributes within a single marker. Within Excel Visualizer there is a single marker per each cell in the spreadsheet, thereby limiting the market to representing a single value. The benefit of multifunctioning visual elements (glyphs) cannot be realized because of the overall metaphor limitation.

**Strict Linkage to Contiguous Ranges**. Excel Visualizer operates on a user-selected range. While this works well with simple uses of a spreadsheet, more complex sheets where variable rows and columns are used, or when pivot tables (where columns and rows can be added interactively by the user) Additionally, blank, hidden and zero-sized rows and columns are simply treated as empty cells.

## 4. EXCEL VISUALIZER USE CASES

Although Excel Visualizer was built as a research application and framework upon which to build and test other spreadsheet-based data visualizations, some users have found practical applications for Excel Visualizer in its current state.

### 4.1 Data Validation

A pre-cursor to Excel Visualizer, which encouraged the development and experimentation of Excel Visualizer was a data cleansing application.

A client worked with large amounts of financial time-series data within spreadsheets. These time series data presented a number of challenges:

- Historic times series data originated from publicly available data feeds and websites of varying quality resulting in occasional data errors and omissions. Using erroneous data in forecasting software would result in a greater margin of error in the forecasts and would therefore reduce the efficiency in capital utilization. Data quality was an issue.
- The firm also generated forecasts of exposure across their holdings over time. Discrepancies in the modeling of different transaction types would be lost as data was aggregated. A discrepancy in values over a few cells may not easily detected by an analyst building a model, nor by inspection of the formulas, but it could be visually obvious.

Prior to using visualization software, the analysts had tried to use charting, but the charting built-in to the spreadsheet proved inadequate for these datasets with 10,000 to 100,000+ data elements. Instead, the analysts created macros to review the data. These macros generated messages and formats based on statistical analysis of the data and the surrounding values. As the amount of reference data increased, the data collection, validation and cleansing step grew to require the work of multiple staff members.

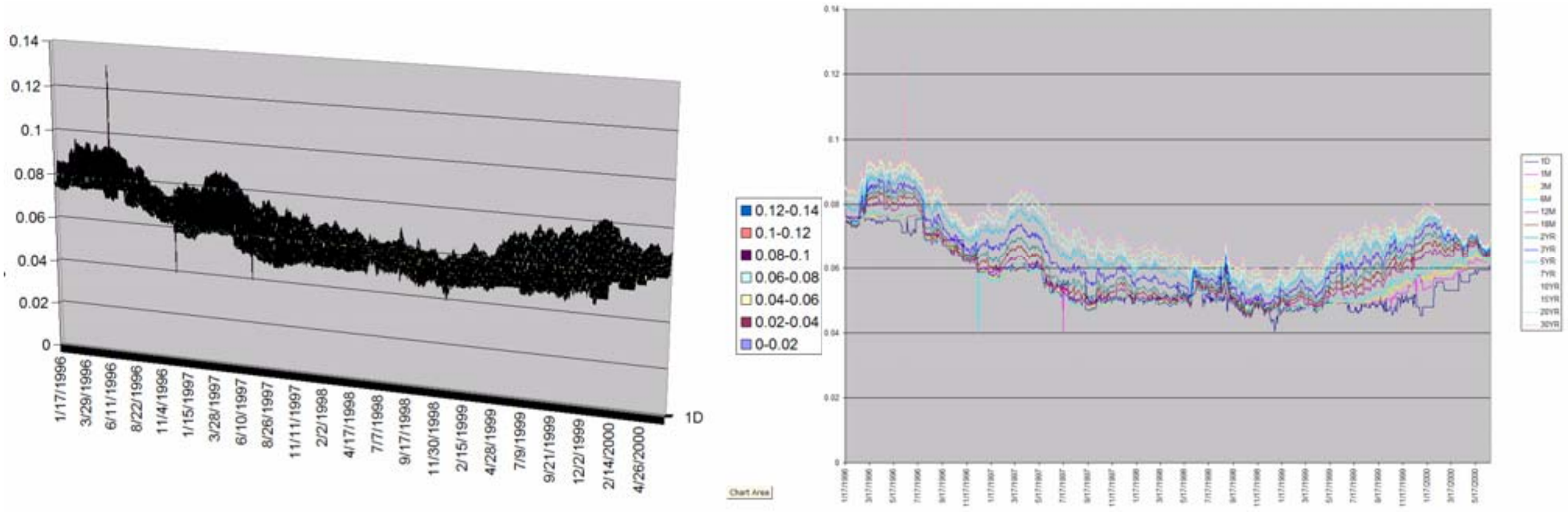

Figure 6. Excel charts of 16 related timeseries (Australian yields) over 1100 days. The 3D chart is not useful. The 2D chart does provide some value, but suffers from occlusion.

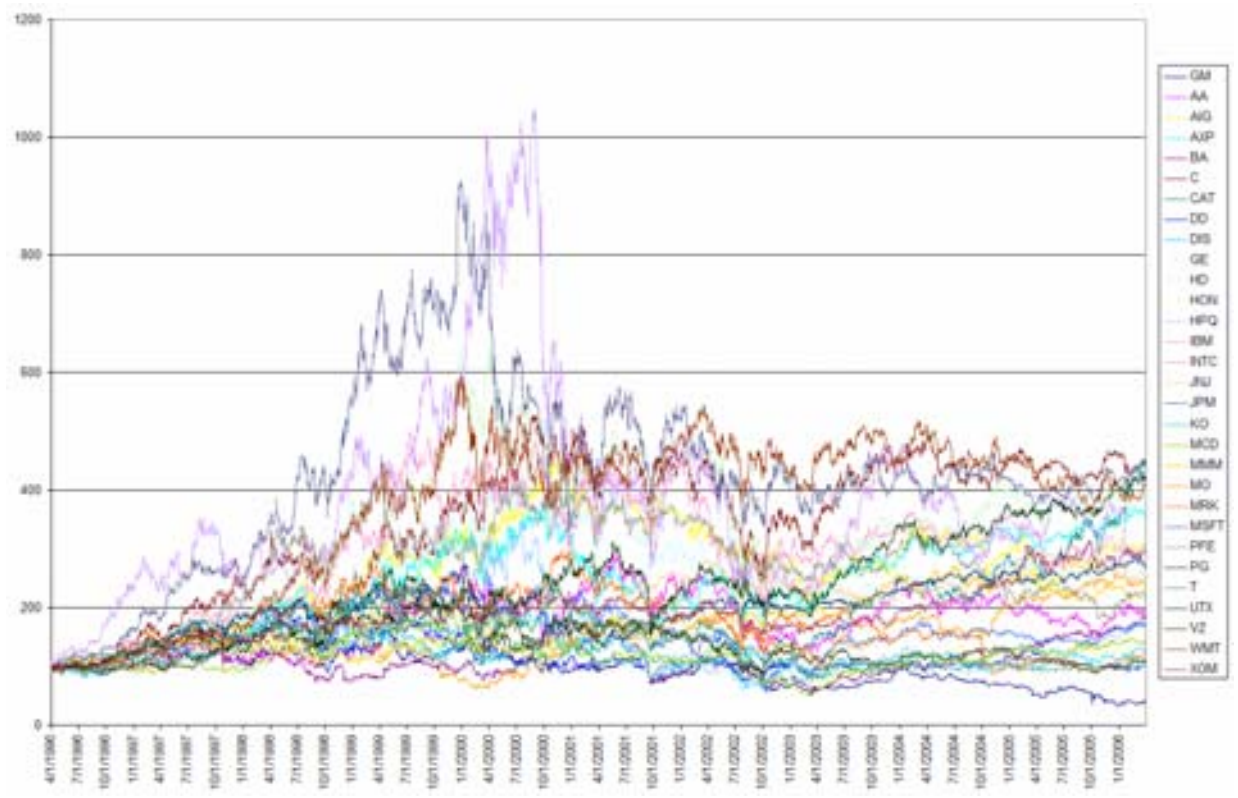

Figure 7. Excel chart viewing 30 timeseries each with 2500+ intervals. Broad patterns are visible, but there is a lot of occlusion in some areas.

Using visualizations of the timeseries data, the analysts are able to see many more cells simultaneously. Outliers are immediately visible. Visual inspection of the outlying bars/surfaces and the immediately surrounding bars, in conjunction with tooltips, provides the analyst with a immediate feedback and the analyst can assess whether the value is within a reasonable range. Exploring further, the analyst can click on the anomalous bar in the visualization to select the corresponding cell in the spreadsheet, providing context (numeric values, row and column information, etc) upon which the viewer could then determine whether to evaluate the data point in more detail (e.g. attempt to validate the value from secondary sources). Finally, by editing the value in the spreadsheet, the analyst can see the immediate update in the visualization display, with immediate visual feedback to determine if the new value “fits”. The visual solution was eventually made available to one hundred analysts.

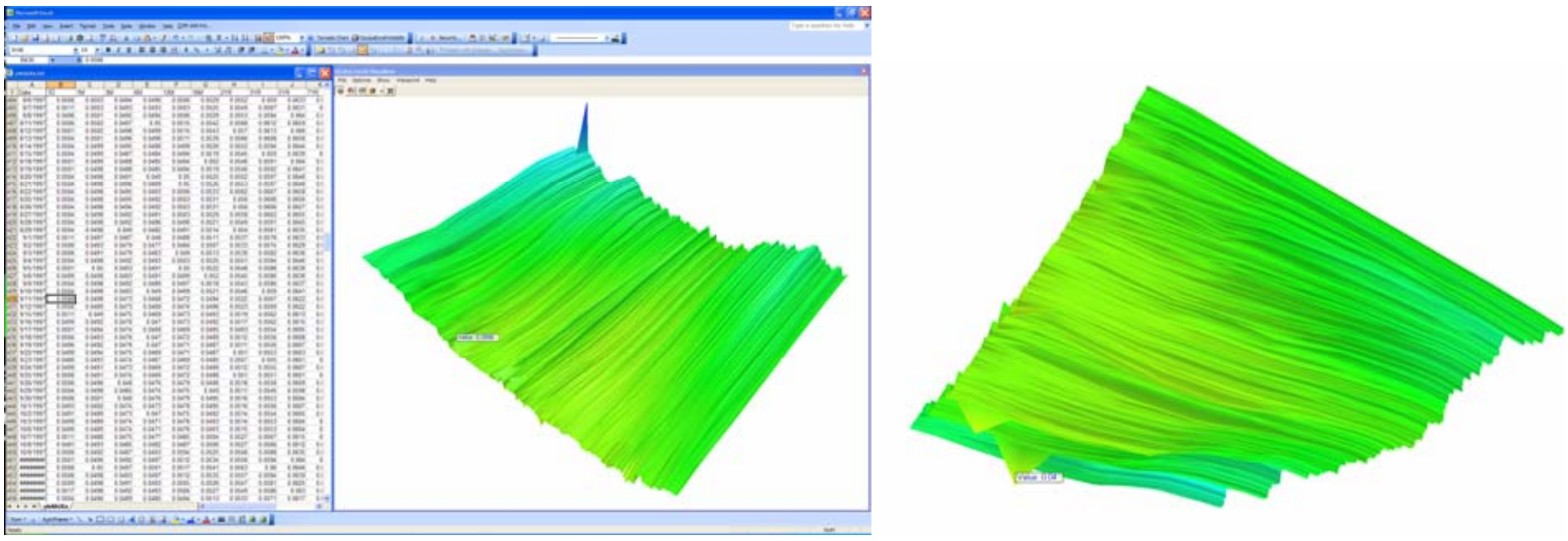

Figure 8. Excel Visualizer viewing 16 timeseries over 1100 days (Australian yields). Fins and tabs are immediately visible indicating anomalies in the data; as are patterns such as ripples, dips and twists, some of which may warrant further inspection. The scene can be easily rotated to different viewpoints to see around occlusions, e.g. the second viewpoint shows anomalies on the underside of the surface.

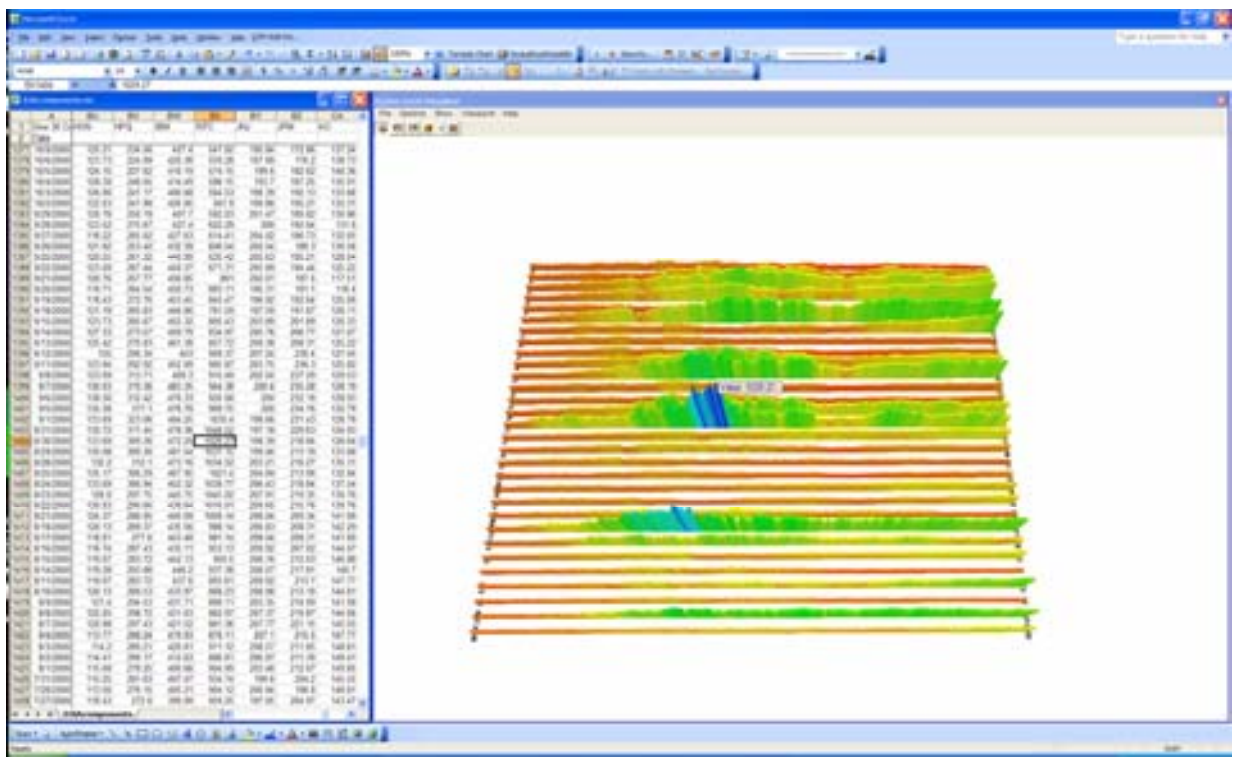

Figure 9. Excel Visualizer viewing 30 timeseries with 2500+ intervals above. Individual timeseries are clearly distinguishable and significant anomalies within a timeseries are readily visible.

With regards to visualizing the results of models, the model builder can view large amounts of data generated by the spreadsheet model and may do further analysis, such as running a back test comparison of modeled values compared to market values. The visualization provides insight into behaviour of the model and may indicate discrepancies in the model. One model builder said: “The visualizations helped us see that there was more basis risk than we were aware of.”

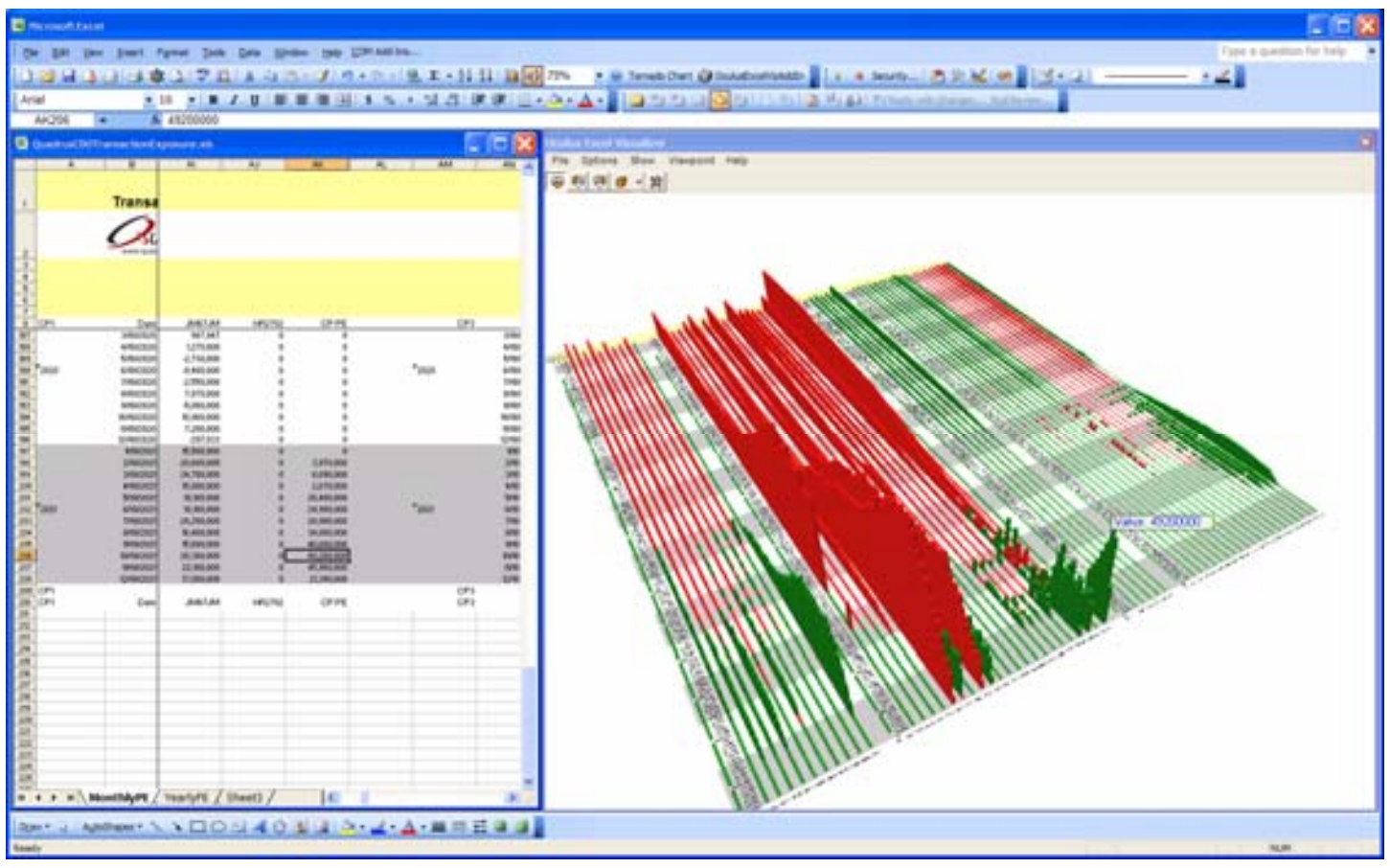

Figure 10. An Excel Visualizer view of a model forecasting exposures for approximately 50 transactions over 200 time intervals. Spikes and other patterns are immediately visible. These patterns may provide insight into the behaviour of the model.

### 4.2 Data Analysis

In another application, Excel Visualizer proved valuable to analysis of event data organized into large, sparse pivot tables (e.g. 20-100 categories per side). It was impossible to see meaningful patterns without reading the value and pivot charts proved to be ineffective. Using Excel Visualizer, the analysts were able to view many more cells at once and see trends and patterns across the data.

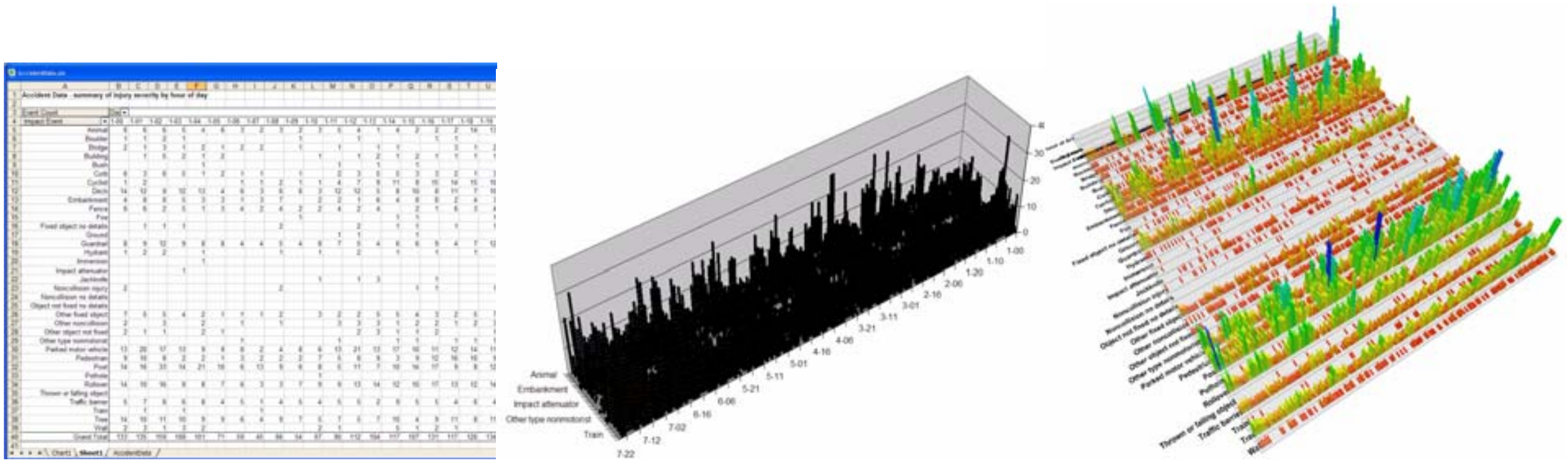

Figure 11. A sample spreadsheet pivot table (36 x 170 cells), shown in a pivot table (left); pivot chart (middle) and in Excel Visualizer (right). In Excel Visualizer, various peaks, strips and cyclical patterns are clearly visible.

As a result of using Excel Visualizer in conjunction with large, flat pivot tables, the analysts were able to reduce the analysis effort to less than one sixth of the previous effort and significantly improve the quality of the analysis.

## CONCLUSIONS

WYSIWYG visualization of spreadsheets is a powerful means to explore the data values in a spreadsheet in context of other values to provide a visual indication of potential outliers or other discrepancies in expected patterns. Using visualization to provide the "big picture" together with the spreadsheet to inspect individual values provides and effective way to review large, data-dense spreadsheets. There are a number of challenges yet to be addressed to make this technique fully effective. In the future, this technique could be generalized and extended to include representations for the calculation network.

## ACKNOWLEDGEMENTS

The data provided in the sample images was created by the author to protect The data viewed in image 7 and 9 are from nasdaq.com. The

Blank Page